# Three-stage Origin of Life as a Result of Directional Darwinian Evolution

## Victor P. Novikov

Nalichnaya 36, 199226
Saint-Petersburg, Russian Federation
vpn@lfp.spb.su

## Abstract

The original hypothesis about Three-stage origin of life (TOL) on the Earth is developed and discussed. The role of the temperature factor in life origin is considered. It is supposed, that three stages of abiogenesis (DNA world, RNA world and the Protein world) consistently followed each other during Darwinian evolution. At the same time, the natural directional selection of the most stable macromolecules and effective catalytic reactions took place. The direction of this selection is related to action of the principle of «Increasing Independence from the Environment» (IIE) and is caused by temperature evolution of the atmosphere of the Earth. The direction of Anagenesis and inevitability of occurrence of genetic mechanisms is discussed.

**Key words:  Origin of Life, Darwinian evolution, Directional selection, Abiogenesis, RNA world, DNA world, IIE principle, TOL hypothesis**

## Introduction

It is known, that there are two counteracting selection processes deferred in Darwinian evolution – stabilizing and directional! The stabilizing selection conserves adaptive properties of organisms, and the directional selection promotes the occurrence of new properties. Eventually the directional selection promotes the progressive evolution of organisms; as a result of this evolution organisms become more complex and improved!

However, the directional selection means also some directional force, and thus, the direction, where the evolution moves to. Only an environment, in which the evolution takes place, can be such directional force, or, more precisely, certain natural changes of environment, that cause corresponding evolutionary changes of organisms. Therefore, of course, when hypotheses about life origin on the Earth is considered, it is necessary to understand, first of all, what evolutionary changes of environment on the Earth promoted progressive biological evolution. Thus, it is important to find out not only a direction of evolution of the environment, but also how it correlated to progressive evolution of organisms.

It is necessary to notice, that the «progressive evolution» term has a little emotional shade and has to be specified. What changes of properties of biological organisms are to be named as progressive? Complexity increase, increase in sizes, new adaptive properties ....? Lot of criteria can be selected; however, the most important character of progressive biological evolution is increasing independence of organisms from the environment! So, for example, at life origin on the Earth the metabolism of organisms was very much depended on environmental conditions, its temperature, pressure, pH and humidity, but afterwards organisms were changing and could already function in various ecological niches.

Gradually live beings left water, have mastered a land, have learnt to fly and were settled from equator to poles of the Earth, have become to huge degree independent of conditions they live in. Therefore, it is possible to assert about organism, that the more independent of the environment it is, the higher step of progressive evolution it takes! The same cardinal principle, undoubtedly, operated at abiogenesis of the first macromolecules on the prebiotic Earth.

But if life origin is not considered as a miracle or unique event, it is necessary to agree that this process is natural. Therefore, of course, similar laws acted throughout all history of the Earth and have caused abiogenesis and biogenesis progressed at the same direction towards progressive evolution.

Thus, the IIE principle (Increasing Independence from the Environment) is a cardinal principle of progressive evolution and its direction. Undoubtedly, IIE should become the basic criterion at studying of the process of life origin on the Earth. And, certainly, it is very important in this regard to understand the relation of the cardinal principle to the direction of evolution of the atmosphere of the Earth and its stages!

## Life origin on the Earth and stages of evolution of the atmosphere of the Earth.

The Earth was formed about 4.5 Ga (billion years ago) and life began on its surface within one billion years. The first live beings on the Earth are thought to be single cell prokaryotes, perhaps evolved from protobionts (organic molecules surrounded by a membrane-like structure) [1]. The oldest ancient fossil microbe-like objects are dated 3.5 Ga (billion) years old, approximately one billion years after the formation of the Earth itself [2] [3], with reliable fossil evidence of the first life found in rocks 3.4 Ga old [4].

Thus, first live beings have arisen most probably at the beginning of the Archean, when the temperature of Earth's crust has decreased and water steam condensed into water. When the Archean began, the Earth's heat flow was much higher than today, and it was still twice the current level at the transition from the Archean to the Proterozoic (2.5 Ga). The extra heat was the result of a mix of remnant heat from planetary accretion, heat from the formation of the Earth's core, and heat produced by radioactive elements.

Cyanobacterial community and archaea [5] were the dominated form of life at the beginning of the Archean eon. However, they, undoubtedly, were the product of long evolution of some earlier elementary organisms. And, certainly, abiogenesis (formation of organic compounds outside of an organism without participation of enzymes) has begun much earlier, than first live organisms have risen! Most likely, it had occurred before the Archean, during Hadean, hundred millions years before the first cells have risen.

The Hadean started with the formation of the Earth about 4.7 Ga ago and ended roughly 3.8 Ga ago. As it is believed, the Earth passed the stage of melting of its external layer at the beginning of its history. The external layer of the planet has been melted to the depth of several tens of kilometers. At that time there was neither hydrosphere, nor oxygen in the atmosphere. An intensive volcanic activity took place on the planet's territory at that time.

At the initial stage of history of the Earth thermodynamic conditions on the Earth's surface reached extreme values: temperature - to 10000 °C and pressure - to $10^9$ Pa. At the middle of the Hadean the temperature of the surface of our planet has decreased to 1000 °C, and in the Proterozoic has decreased to 40 °C. In the Cretaceous period (about 100 Ma ago) the temperature of the surface in average was 22 °C, and at present is about 15 °C [6].

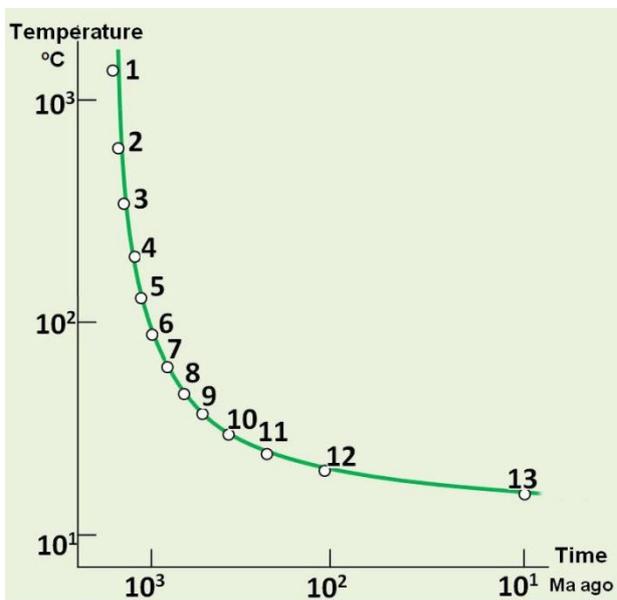

Billions years of temperature evolution have caused occurrence of conditions for life origin on our planet. Then further biological evolution of live beings to their higher, more improved organization to huge degree correlated to temperature evolution of the Earth. Thus, the temperature factor, undoubtedly, was the main directing force of Anagenesis.

**Figure 1. Conjugation of Anagenesis with temperature evolution of the surface of the Earth.**
*1 — synthesis of simple organic molecules, 2 — optimum of non-enzymatic DNA synthesis, 3 — optimum of non-enzymatic RNA synthesis, 4 — optimum of synthesis of polypeptides, 5 — formation of oceans, 6 — occurrence of Prokaryotes, 7 — first aerobes, 8 — Eukaryote, 9 — occurrence of multi-cells, 10 — occurrence of vertebral, 11 — first reptiles, 12 — occurrence of mammals, 13 — beginning of hominization of primates.*

According to the presented Figure 1, unidirectional change of temperature of the environment has provided action of natural selection towards reduction of temperature dependence of organisms from thermodynamic conditions of the environment [7]. Thus, the cardinal IIE principle operated during both abiogenesis and biogenesis.

It is necessary to notice, that the conditions could be different throughout all history of the Earth. Very warm periods have happened; they were followed by freezing periods. However, a singular tendency of change of average temperature for millions years took place. It was gradual cooling of the surface of the planet. Freezing periods are recognized by corresponding burst of speciation, and warm epochs — by static periods, when the intensity of Anagenesis decreased. Consecutive decrease in temperature of the surface of the planet from extreme to present values promoted selection of such molecular mechanisms, and then species, which, to some degree, by this way or another, could support their independence from the temperature of the environment.

Synthesis and primary selection of the most stable biopolymers were going on at that stage of history of the Earth, when the temperature of its surface was still considerably above the water boiling point; it means, that abiogenesis took place in waterless (dehydration) conditions. Heat energy had the prior role as the energy source in chemical evolution on the Earth, when the first simple organic compounds have risen [8]. Apparently, increase of independence of metabolism from the temperature of environment during abiogenesis is the main exponent of IIE. Hence, the numerical value of IIE can be characterized by the formula $R = \frac{T_{out} - T_{in}}{T_{out}}$ , where $T_{out}$ — the temperature of environment, $T_{in}$ —the optimum temperature for the metabolism, R — the indicator showing the numerical value of the IIE. It is possible to allocate three stages of abiogenesis, closely related to temperature evolution of the Earth's surface (Fig. 2).

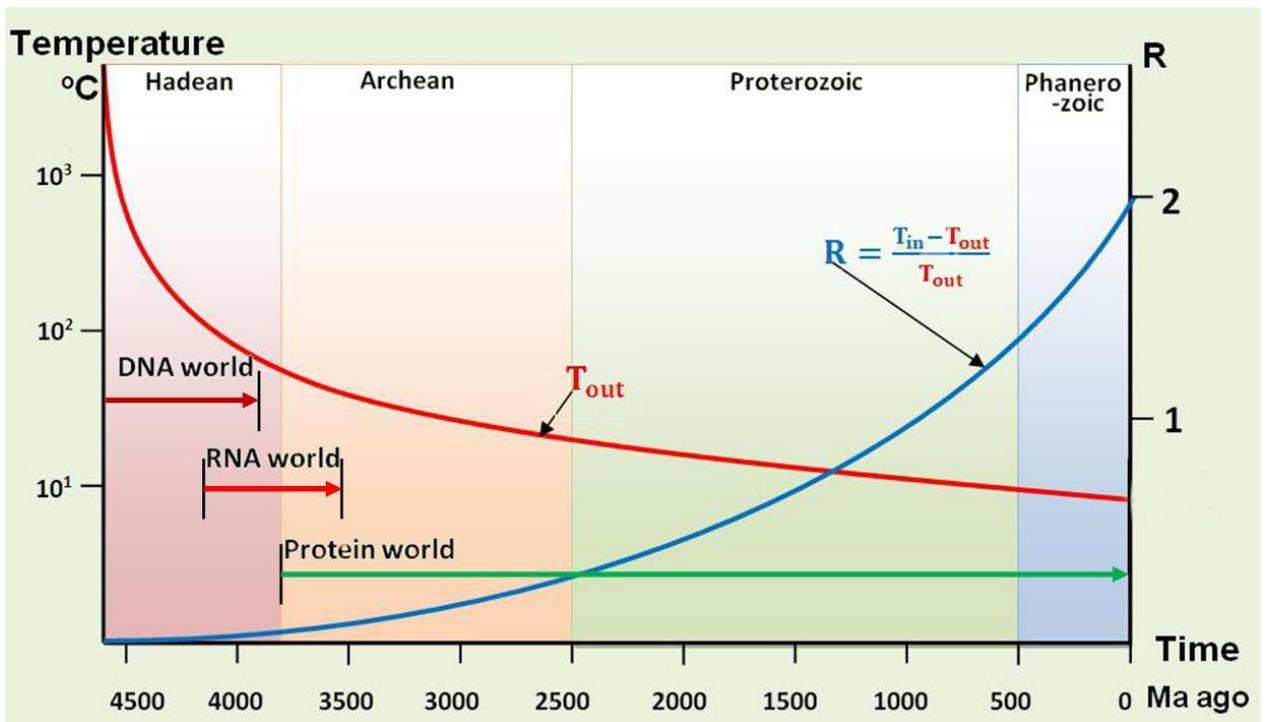

**Figure 2. Stages of abiogenesis and temperature evolution of the Earth**
*$T_{out}$ — ambient temperature, $T_{in}$ — temperature of metabolism, R — shows the numerical value of IIE*

As it is shown in the fig. 3, DNA world takes almost the entire Hadean. RNA world arises in the second half of the Hadean (rather, as DNA+RNA world) and is the evolutionary bridge between waterless DNA world and water Protein world.

Chemical abiogenous evolution took place in waterless conditions in the range of temperatures from 6000°C (a temperature limit of stability of the elementary diatomic compounds OH, $H_2$, etc.) to 100°C (temperature of condensation of steams of water). During this evolution, caused by gradual decrease of the average temperature of the surface of the Earth, DNA, RNA and amino acids, consistently (in various temperature intervals) appeared. And from 400 $^{o}$C onward, there were autocatalytic microsystems [9] originated from them. The microsystems could successfully pass a power barrier of the reaction of the replication. As it is known, in average the energy of activation of non- catalytic reactions makes 30-45 kcal, heterogeneous catalytic - 15-30 kcal, and enzymatic - 8-12 kcal [10].

## DNA world - genesis of the first macromolecules

Besides phosphates there were only reduced forms of chemical compounds (for example, $CH_4$, $CO$, $H_2$, $NH_3$, $H_2S$) in sufficient quantity available in the heated atmosphere of the Earth during the Hadean. And, certainly, the ancient atmosphere didn't contain any molecular oxygen.

Thus, during the Hadean the abiogenesis took place at extreme conditions, which were very much differed from living conditions of organisms at present. High temperature, absence of water, acid steams and an intensive ultraviolet made the abiogenous synthesis of proteins impossible.

Therefore, at the Hadean the abiogenesis of only such macromolecules as nucleonic acids, that can be synthesized only in waterless conditions, was possible. As it is known, water is required for synthesis of proteins! Whereas, for example, ester links in RNA between ribose and phosphoric acid are prone to hydrolysis.

Ultra-violet radiation of the Sun, heat of volcanic processes, ionizing radiations of radioactive decay and electric discharges at that time could act as an energy source, initializing the synthesis of organic molecules. It is also possible, that Oxidation-reduction processes between volcanic gases (reducer) and partially oxidizing sulphide minerals, for example a pyrite ($FeS_2$), could be a source of energy, needed for creation of bio-molecules.

Synthesis of some organic molecules, apparently, has started already at 1000 °C, right after thermal dissipation of a considerable part of hydrogen from primary atmosphere of the Earth [12] has occurred, as it interferes, for example, with formation of nitrous bases [13]. Predecessors of pyrimidines – cyanogen-acetylene and urea, elementary sugar – formaldehyde [14], the predecessor of adenine - cyanic hydrogen [15], nitriles - predecessors of amino acids [16] and other intermediate products of synthesis of biologically valuable molecules were formed in significant volume at high temperature and pressure, in the atmosphere, containing carbonic gas, nitrogen, methane, ammonia and water steams. Then, or simultaneously, nitrogenous bases, amino acids and carbohydrates rose.

Not only synthesis, but also continuous disintegration of compounds took place. The most stable compounds were accumulated, for example, cycling forms, which could serve as a matrix for the subsequent compounds. Besides that, surfaces of minerals [17] could be an initial matrix. Also, the ability of DNA to form secondary and tertiary structures also increased their stability as it protected the most reactionary centers of a molecule.

The maximum accumulation of more stable compounds took place at certain temperature or in the range of temperatures. In these high-temperature conditions natural selection of molecules by their stability and their accumulation in liquids, for example of hydrocarbonic nature, as waters then didn't exist yet was carried out. Thus, Darwinian selection operated even at level of the first macromolecules.

DNAs were the first compounds that had their optimum during temperature evolution, because they are the most stable among polymers. It was the first stage of chemical evolution of organic connections. During this period a selection of the most stable nucleotide sequences took place. It is confirmed by the fact, that evolutionary earlier organisms had nucleic acids of GC type [18] as DNA of this type are more resistant to high temperature and the GC pairs are much stronger, than AT pairs, stabilize interactions between series of the bases, that is an important advantage in processes of replication [19].

It should be noted that macromolecules of DNA are not only very stable, but also possess catalytic activity [20]. Already 30 years ago it is revealed, that DNA, as well as enzymes, can serve as catalysts in different biochemical reactions. Ability of Deoxiribozymes to decompose, to ligate and to phosphorylate molecules of DNA [21], also to methylate the porphyrins is described. In the Archean, apparently, homochirality has occurred. The directional selection of D-DNA took place, as replication of D-DNA under the influence of ultra-violet radiation [22] was easier carried out.

Thus, at abiogenesis of series of DNA during millions of years billions of molecules went through Darwinian selection. The most stable and capable to self-replication molecules were selected by natural way. Selection of molecules of DNA and their components took place in the world where volcanoes erupted frequently, thunder-storms and earthquakes happened. In primitive atmosphere all was constantly mixed up, gurgled and was stirred up. Sometimes some components were collected in secluded places or in crevices of rocks, were transferred by air, etc.

Certainly, they don't speak about occurrence of any genetic mechanisms at this stage of abiogenesis. The nature only developed by selection the stability of molecules of DNA, their resistibility to disintegration under influence of ultra-violet radiation and ability to replication. There were a selection of the most stable variants of DNA with various sequences of the nitrogenous bases, a selection of the suitable nitrogenous bases for primary sequence and a selection at the level of secondary structure in DNA world.

Besides, the most economic ways of synthesis of these macromolecules were chosen. Therefore, this selection wasn't casual, and occurred according to the cardinal IIE principle of Anagenesis. Thus, the independence from DNA from environment by means of selection of more stable variants of macromolecules increased. Simultaneously replication capabilities of DNA by selection of suitable catalytic reactions became stronger.

This Darwinian selection of components of DNA occurred millions years in parallel and simultaneously in various places of the Earth. Billions of macromolecules of DNA passed selection in the same direction. It accelerated evolution and led to identical results. In this regard DNA of all beings on the planet consists of the similar nitrogenous bases. After all, physical and chemical laws operate similarly everywhere. It is possible, that DNA of the majority of beings of the Universe also consists of similar components. We aren't alone in the Universe. Everywhere the planets capable to generate the life pass evolution stages similar to the Earth's. And everywhere evolution of atmospheres of planets passes through the stage of high temperature and the subsequent cooling. Therefore, RNA world inevitably comes after DNA world at decrease in temperature of a surface of a planet.

## RNA world - evolution of genetic machinery

At this second stage of chemical evolution the power barrier of reaction of DNA replication on DNA matrix has considerably increased. As a result of decrease in average temperature of the surface of the Earth the temperature optimum of DNA synthesis on DNA matrixes has passed and the optimum of RNA synthesis has begun.

Strengthening of RNA synthesis in these conditions facilitated DNA formation as this process began in two stages: in the beginning RNA on DNA-matrix, and then- DNA on RNA-matrix. Hence, during temperature evolution an autocatalytic system was created from two components: DNA↔RNA which were in the aggregated condition in unpolar liquids, forming microsystems of coacervates' type. These structures were the bubbles, which covers were consisted of lipids, and inside there were DNA molecules carrying out the replication.

Experiences with artificial protocells have shown that such bubbles spontaneously form new bubbles, a new portion of lipids being added. So all it occurs automatically and without any influence of DNA [23]. Hence, in the past the protocells could occur in the RNA-world and reproduct without enzymes at presence of ribozymes [24].

Many natural ribozymes catalyze either the hydrolysis of one of their own phosphodiester bonds (self-cleaving ribozymes), or the hydrolysis of bonds in other RNAs. Investigators studying the origin of life have produced in the laboratory ribozymes, which are capable to catalyzing their own synthesis under very specific conditions, such as an RNA polymerase ribozyme [25]. Laboratory samples, however, have shown low catalytic ability: they are in time to collect series of not more than 14 nucleotides for 24 hours and afterwards they decay due to hydrolysis of phosphodiephyr bonds. However, it can be explained by water presence. After all, in the beginning of RNA epoch these ribozymes operated in waterless conditions and were more active.

It is necessary to note, that DNA and RNA prebiotic molecules acquired the capacity to store the genetic information for the first time exactly in RNA world. The feedback DNA↔RNA has arisen. A selection of the most catalytically active ribozymes and corresponding DNA series took place. Thus, the Darwinian selection of the most catalytically active ribozymes was processed and the corresponding genetic information was fixed in DNA.

The subsequent gradual decrease in ambient temperature has increased the power barrier of reaction of DNA synthesis on RNA-matrix and has led to the following stage of chemical evolution. The temperature optimum of abiogenous, non-enzyme synthesis of proteins from amino acids has started. Researches show, that in the beginning protein synthesis was carried out by means of only small RNA molecule, capable to catalyze the reaction of composition of two amino acids [26]. All other structural blocks of ribosome in the course of evolution were consistently added to a protoribosome, without breaking its structure and gradually increasing efficiency of its work.

At first it was mostly nonspecific process which, however, was going on much faster and more effectively in presence of the polynucleotides, forming stable complexes with proteins. By the way, these complexes aren't hydrolyzed even in boiling water. This fact can be related to RNA presence in membranes of modern cells [27], where these polynucleotides also form complexes with proteins.

It is interesting to notice that, apparently, at this stage of chemical evolution there was an infringement of symmetry of distribution of optical isomers of amino acids in prebiological systems. According to some data [28], left- side amino acids have lower power barrier of reaction of polymerization of amino acids, than right-side isomers. Hence, during evolution there was a selection of the left isomers most convenient in power regard.

At later evolutionary stage synthesis of proteins passed through a stage of aminoaciladenilate formation. It is shown in modeling experiences, that thermal condensation of aminoaciladenilate mix is processed very quickly [29] and formed proteins, have various, though weak, catalytic activity [30] in particular they catalyze formation of internucleotide links. These characteristics of proteins, along with their ability to form microspheres, were the original chemical revolution and promoted enzyme overcoming of the power barrier of replication reaction, and also transition of these already phase-isolated systems in the water environment.

## Protein world - origin and evolution of cells

The Protein world, apparently, has occurred after the temperature on the planet's surface has decreased to 1OO °C and water began to condense. Therefore, the first cellular organisms could arise in the hot water environment nearby 3.7 Ga years back. It is proved by the data that Archaea (phylogenetically most ancient microorganisms) are the obligate thermophiles [31].

Abiogenous evolution in the direction of origin of the first autocatalytic phase-isolated systems has been caused and closely interfaced to temperature evolution of the planet's surface. By the way, thereby the known contradiction between high-temperature dehydrotating conditions of experimental synthesis of biopolymers [32] and existence of modern systems of biosynthesis in the water environment at room temperature is eliminated. At evolutionary approach it becomes clear, that it is not the contradiction of the two parties, but two consecutive stages of life origin.

The occurrence of water at 100 °C on the planet has caused the transformation of abiogenous evolution into biological one and became an important component of natural selection of the phase-isolated microsystems (protocells) as the problem of their protection against hydrolysis has arisen. As a matter of fact, already phase isolation of first protocells, probably, was the first element of their protection against destructive influence of the water environment, as an area free of water molecules could be in these microsystems [33]. Other factors, interfering hydrolysis, could be spiralization of biopolymers and formation of tertiary structures.

The occurrence of the Protein world can be named as the revolution in an abiogenesis. Application of proteins made it possible to increase catalytic capabilities of RNA molecules by many times. The originated enzymes possessed not only high speed of catalysis, but also have sharply increased the substrate specificity of biochemical reactions, their selectivity. On the one hand, the origin of a cellular structure has many times increased the degree of independence from biochemical reactions from the environment. On the other hand, the increased complexity of biological system has increased the degree of influence of mutations on a phenotype and on efficiency of Darwinian selection.

The birth of modern genetic code was the result of a struggle of these two opposite tendencies. In the beginning it was only some casual affinity (mutual attraction) between some short RNA and amino acids. Then during evolution the settlement of genetically useful variants of primary structure of DNA took place. It is considered, that some similarity of genetic code between all organisms specifies the presence of the last common ancestor (LCA), which caused all species known by science that have extended in the course of evolution. For example, Haeckel dreamed to find such organism, which could be placed in the basis of a tree, and even once has informed, that it is found. However, LCA is not found till now and, hardly it will happen in the future. After all actually the similarity of a genetic code of various species is caused not by the same way of origin, but by evolution in the same direction. In the beginning, abiogenic directed evolution of DNA and RNA on preobiotik Earth took place. Then the first primitive cells Anagenesis occurred towards increasing the independence from the external environment.

The directional natural selection of more power convenient systems of biosynthesis of amino acids also confirms it. The data of statistically authentic correlation between average values of amino acid presence in proteins and ATF input at synthesis of amino acids from glucose also show its evidence. ATF input to amino acid biosynthesis being increased its contents in proteins decreases [34]. Especially considerable ATF input is demanded for synthesis of the majority of irreplaceable amino acids.

Besides, a natural selection of L-amino acids and D-sugars took place during biological evolution, because, as calculations show, the energy of their basic state lower, than the energy of their mirror antipodes. It is supposed [35] that it is caused by the contribution of weak interactions to interaction of electrons with the core of the asymmetric center of the hyral molecule. Hence, a selection of not only organisms of more effective systems of replication of polynucleotides, but also most energetically convenient substrata of biochemical reactions of translation system took place during anagenesis.

Certainly, Darwinian evolution (both abiogenous, and biogene) wasn't straight. The deviations, the periods of stasis and the stages of skips of speciation took place. However, the general tendency remained the same. The purposeful anagenesis occurred under the influence of the directional Darwinian selection. As a result of action of the cardinal IIE principle cellular organisms acquired the new, more improved biochemical mechanisms. The oxygenation of the atmosphere has begun approximately 2.4 Ga years back [36] and promoted it.

Then Eukaryotes have risen. The earliest proof of their occurrence is 1.85 Ga years old, though, probably, they have risen even earlier. The diversification of Eukaryotes was accelerated, when they have started to use oxygen in metabolism. Later, nearby 1.7 Ga years ago, multicell organisms with the differentiated cells for performance of specialized functions [37] have risen. First seaweed rose approximately 1.2 Ga years ago, and first higher plants —about 450 Ma years ago [38]. Invertebrate animals have risen in the Ediacaran period, and the vertebral have risen about 525 Ma years ago during the Kambrian peak of speciation [39].

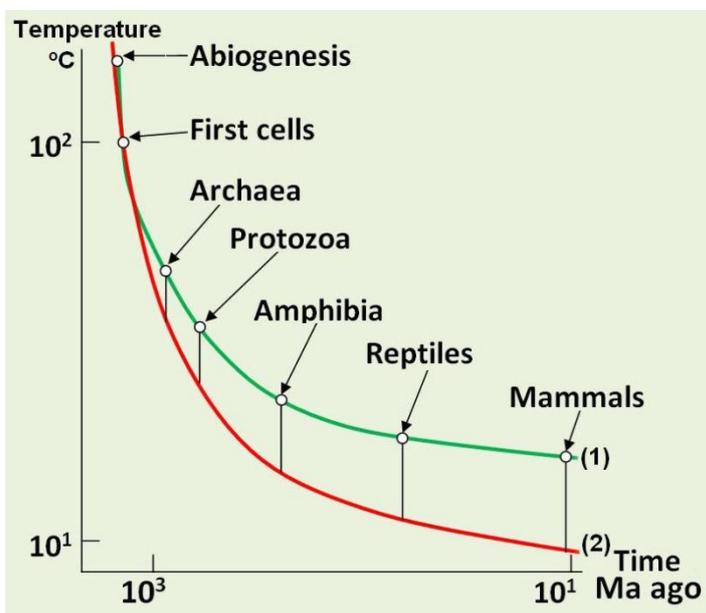

Consecutive decrease in temperature of the planet's surface promoted the selection of such molecular mechanisms, and then species, which to some degree could that way or another support relative independence of their functioning from temperature of the environment. Biogenesis of internal cellular processes has started after thermo-chemical abiogenesis. Then anagenesis of multicell organisms has come. Thus, the higher step of a philological ladder the specie is, the wider temperature range it can live in (fig. 3).

**Figure 3. Temperature gradient, accompanying biological evolution**
*Where: (1) — optimum of the internal medium of an organism, (2) — average temperature of the surface of the Earth*

It is shown on fig. 3, how the difference between temperatures of the environment and an organism increased during evolution, i.e. the temperature gradient between them increased. It demanded more and more reliable heat producing and temperature homeostasis systems of live beings. And though the temperature optimum of functioning of biological systems all the same decreased during anagenesis, at the same time, there was a continuous natural selection of organisms having more intensive power metabolism, more independent from temperature conditions of the environment. In this regard the difference in temperatures of the environment and an organism (a temperature gradient accompanying biological evolution) grew as well.

Temperature evolution of the Earth, especially its initial stages, was defined by action of the second law of thermodynamics about irreversibility of thermal processes in the nature. So as biological evolution correlates with temperature, it means, that origin and progressive evolution of live organisms also is not casual, but strictly determined process which direction is defined by the same second principle of thermodynamics.

# Conclusion

At present the big success in the research of life origin is reached. However, it is difficult to expect that in the near future there will be fast and final success on this way. Three-stage Origin of Life (TOL) hypothesis is a fundamentally new perspective in the research of Origin and Evolution of Life. It represents a theoretical scheme of an abiogenesis and its details are still to be specified. However, now it is already proved by many facts and is the most harmonious and logical. Its integration to Hadean's DNA world hypothesis and IIE principle make it unique and consistent. Waterless Hadean has given naturalness to an abiogenesis, and IIE has combined the abiotic evolution with the biotic evolution.

IIE principle has defined the direction of natural Darwinian selection and has made it related to the temperature evolution of the atmosphere of the Earth. The directivity of the selection made it possible to reject unstable macromolecules, inefficient reactions and unadapted biological species. Therefore TOL hypothesis makes it possible to reduce the time necessary for life origin to a reasonable period and is an excellent alternative to current mainstream evolutionary theories about casual or sudden origin of life.

TOL hypothesis by means of IIE principle helps to understand inevitability of life origin and anagenesis direction. It explains conformity of genetic mechanisms' origin to natural law and characteristics of similarity in genetics of various organisms. IIE principle is the basic characteristic of anagenesis and somewhat its purpose.

Temperature evolution of the atmosphere of the Earth has defined a directivity of anagenesis to reduction of dependence of biological systems from temperature environmental conditions. This evolution due to the irregularity, promoted the intermittent character of anagenesis of organisms. Three-stage origin of life is a natural process which is caused by thermodynamic evolution of the environment. In this sense a live organism can be characterized as a natural object, which had ancestors passed the directional selection under the influence of the second principle of thermodynamics during hundreds millions years.

TOL hypothesis in wide sense transfers the Darwinian selection to abiogenesis. The natural directional Darwinian selection operates at abiogenesis, as well as at biogenesis. In the first case it selects the most stable macromolecules, in the second – the organisms most independent from the environment. However, in the given context the stability and the independence are somewhat synonyms.

Certainly, it is impossible to find the single determinant proof of Darwinian selection process at abiogenesis. The same is regarded to the final experiment which would result in live cell sudden origin in boiling and gurgling mix of simple inorganic substances. Actually the natural abiogenesis processed during hundreds millions years, in parallel in millions natural reactors, at very low activity of natural catalysts.

It should be noted the temperature and chemical conditions of the atmosphere and of Earth's surface changed many times during the Hadean and the Archean. Some macromolecules and their components were accumulated; others decomposed, then again entered the reactions and so on repeatedly. The abiogenesis, as well as any biological evolution wasn't straight linear process and was frequently branched out, interrupted, receded, was slowed down and accelerated. Process of life origin was very long, tortuous and contradicted. In this sense the abiogenesis consists of a range of casual events, casual reactions in accidental conditions. But evolutionary progress is not a random process. The directional Darwinian selection caused by temperature evolution of the atmosphere, provides anagenesis with appropriateness and acceleration.

Research of stage-by-stage abiogenous DNA synthesis from simple inorganic substances with assistance of natural catalysts would be one of the basic experimental proofs of justice of TOL hypothesis. It would be important to investigate the way how to obtain the directional selection of stable macromolecules and effective catalytic reactions of abiogenesis by changing the temperature of the environment and, probably, some other conditions.

Other interesting direction of research is to produce a primitive protocell from DNA, RNA and lipids in waterless environment (and then in the aquatic environment) so that it would be capable for DNA replication and reproduction.

Of course, still a long way of theoretical and experimental researches to solve the problem of life origin is expected.

Thus, perhaps, the TOL hypothesis will make a significant contribution to a fundamental understanding that life origin occurred as a result of Darwinian evolution by natural directional selection!